\documentclass[aps,prd,preprint,amsmath,amssymb,nofootinbib]{revtex4-1}

\usepackage{bm}

\begin{document}

\title{Kinetic theory and evolution of 
cosmological fluctuations with neutrino number asymmetry}

\author{Manuel Valle}
\email[]{manuel.valle@ehu.es}
\affiliation{Departamento de F\'\i sica Te\'orica, 
Universidad del Pa\'is Vasco UPV/EHU, 
Apartado 644,  48080 Bilbao, Spain}

\date{\today}

\begin{abstract}
We derive a kinetic equation for chiral matter at non-zero chemical potential that governs the response of the parity odd part 
of the distribution function to perturbations of the Robertson-Walker metric.  
The derivation is based  on a recent evaluation  
of the gravitational polarization tensor at non-zero chemical potential. 
We also provide the equations for gravity waves that follow from the  anisotropic stress tensor describing the lepton asymmetry. 
These equations  can be used to assess the effects that a non-zero neutrino  chemical potential would have on the 
evolution  of cosmological perturbations.  
\end{abstract}

\pacs{98.80.Cq, 11.10.Wx, 11.30.Rd}


\maketitle

The implications of the existence of  gauge and gravitational anomalies on relativistic hydrodynamics 
are being systematically explored nowadays (see e.g. Refs.~\cite{Landsteiner:2012kd, Jensen:2012kj} and references therein). 
Most of the work  has focused on the new  odd susceptibilities 
and transport coefficients  related to the parity anomaly, and  it has been realized that some 
of these quantities may be obtained from the variation
of the equilibrium partition function in the presence of 
a time independent background of the metric and gauge fields~\cite{Banerjee:2012iz,Jensen:2012jh}. 
On the other hand, the study of  time-dependent processes requires  the
evaluation of the  appropriate Green's functions at non-zero frequency or, alternatively, the  
use of Boltzmann equations describing the evolution of distribution functions in momentum space.   
By considering the response to a background electromagnetic field, 
the authors of Ref.~\cite{Son:2012zy} have obtained a kinetic equation including the effects of 
triangle anomalies  and have also discussed the interplay  between the kinetic and field theoretical approaches (see also  Ref.~\cite{Gao:2012ix}). 
As their analysis does not  include the gravitational response, in order  to complete the description it  would be  necessary to 
consider the gravitational correlation function between energy-momentum tensors.    

The study of the thermal gravitational correlation function at non-zero frequency was performed 
some time ago by  Rebhan~\cite{Rebhan:1990yr}. At very small momenta $Q^\nu=(q^0, \bm{q})$ compared to  the temperature,  
he showed that this quantity is proportional to the thermal energy density and   has an universal tensorial structure that 
obeys the gravitational Ward identities for diffeomorphism and conformal transformations~\cite{Rebhan:1990yr}. 
Later in Refs.~\cite{Kraemmer,Rebhan:1994zw}  these field theoretical results were 
used to work out the evolution of cosmological perturbations, showing that 
they provide an equivalent description to that obtained from the kinetic approach based on the  Vlasov equation.  
More recently,  the authors of Ref.~\cite{Manes:2012hf} have studied the parity violating part 
of the gravitational response of an ideal gas of Weyl fermions at non-zero chemical potential $\mu$.
This next-to-leading contribution, which also satisfies the Ward identities,  
is simply proportional to the net number density of chiral fermions. 
At zero frequency its form may be used to determine the modifications of  the 
constitutive relations of hydrodynamics that give rise to macroscopic parity violating  effects, such as the chiral vortical effect. 
But, as far as we know, the implications of the parity odd contributions for  time-dependent 
gravitational perturbations have not been much explored.  
Previous studies have only focused on the effects of lepton asymmetry related to the dependence of the anisotropic inertia on 
even powers of the chemical potential~\cite{Ichiki:2006rn}, or at most, have  
introduced effective interactions generating cosmological birefringence~\cite{Geng:2007va}.     

In this work we will use  the knowledge of the parity odd correlation function, denoted 
by $\Pi_\epsilon^{\mu \nu \, \rho \sigma}(q^0, \bm{q})$,     
to derive the Botzmann equation that governs the evolution of the 
$\mu$-dependent part of the chiral fermion distribution. 
The kinetic equation thus obtained turns out to be surprisingly simple. 
It includes a source term proportional to  the chemical potential. 
In this way, we complete  the treatment in Ref.~\cite{Son:2012zy}. 
Once adapted  to the Robertson-Walker metric,  these results could be used to 
assess the effects of neutrino asymmetry in the evolution of cosmological perturbations. 
 
We first present the results of the thermal field theoretical calculation of 
$\Pi_\epsilon^{\mu \nu \, \rho \sigma}(q^0, \bm{q})$ in flat space-time. 
The amount of the neutrino asymmetry for a given species is described by  the degeneracy parameter, defined by 
the ratio of the chemical potential to the temperature $\xi_\nu = \mu_\nu^{(0)}/T_0$, which is assumed to be small  $|\xi_\nu|  \ll 1$. 
In terms of the Fermi-Dirac distribution functions
\begin{equation}
\label{fermi}
 \overline{n}_\pm(p) \equiv  \frac{1}{(2\pi)^3} 
 \left[ \exp \Bigl(\frac{p}{T_0} \mp \xi_\nu \Bigr) + 1\right] ^{-1} , 
\end{equation}
the net unperturbed neutrino number density reads 
\begin{equation}
\overline{n}_{\nu - \overline{\nu}}\equiv
 \int_0^\infty 4 \pi p^2 \bigl(\overline{n}_+(p) - \overline{n}_-(p) \bigr) dp  = 
 \frac{T_0^3}{6} \left( \xi_\nu + \frac{\xi_\nu^3}{\pi^2} \right)  \approx \frac{T_0^3 \xi_\nu}{6} . 
\end{equation}
Here $T_0$ is any arbitrary  reference temperature, that in the cosmological setting will be related to  the equilibrium temperature 
at the present time $\overline{T}(t_0)$ through $T_0 = \overline{T}(t_0) a(t_0)$,  
with  $a(t)$ the Robertson-Walker scale factor. 
The Fourier components of the perturbations to the energy-momentum tensor  are connected with metric perturbations  
$h_{\mu \nu}(t, \bm{x}) = g_{\mu \nu}(t, \bm{x}) - \eta_{\mu \nu}$~\footnote{$\eta_{\mu \nu}   =\text{diag}(-1,1,1,1)$, 
$\epsilon^{0 1 2 3}=1$. } 
by 
\begin{equation}
\label{eq:em}
 \delta\langle T^{\mu\nu}(Q)\rangle=-\frac{1}{2} \Pi^{\mu \nu \, \rho \sigma} (q^0, \bm{q}) h_{\rho \sigma}(Q) , 
\end{equation}
where the retarded graviton self-energy has been  defined by 
\begin{equation}\label{correl}
\Pi^{\mu \nu \, \rho \sigma}(x-y) \equiv -i\, \Theta(x^0-y^0) \left\langle \bigl[T^{\mu \nu}(x), T^{\rho \sigma}(y)\bigr ] \right\rangle
-2\left\langle \left. \frac{\delta \bigl(\sqrt{-g(x)}T^{\mu \nu}(x)\bigr)}{\delta g_{\rho \sigma}(y)}\right|_{g=\eta} \right\rangle .  
\end{equation}
The calculation from thermal field theory shows that 
 the thermal part of this response function
receives a parity violating contribution  proportional to the totally antisymmetric symbol $\epsilon$. 
This contribution is tied to the  helicity of the equilibrium thermal state~\cite{Loganayagam:2012pz}, and   
 for $Q \ll |\mu_\nu|, T$, it is suppressed by a factor $\xi_\nu Q/T$ with respect 
to the  leading-order temperature contribution proportional to the energy density 
$\overline{\rho}_{\nu + \overline{\nu}} \approx 7 \pi^2 T^4/120$.
 Because the one-point function $\langle T^{\rho \sigma}\rangle$ does not have any odd-parity contribution, 
 the graviton self-energy tensor verifies the Ward identity  $Q_\mu \Pi^{\mu \nu \, \rho \sigma} (Q) = 0$.  
Its  explicit form is given by  
\begin{equation}\label{proy} 
\begin{split}
\Pi^{\mu \nu \, \rho \sigma}(q^0, \bm{q})&=
i c_{V}(q^0, q) \frac{Q^2}{(u \cdot Q)^2 + Q^2} u_\alpha Q_\beta \bigl[ \epsilon^{\alpha \beta \mu \rho}
    P_V^{\nu \sigma}  + 
    \epsilon^{\alpha \beta \nu \rho}
    P_V^{\mu \sigma}  +   ( \rho \leftrightarrow \sigma ) \bigr]    \\ 
    & \quad   + 
i c_{T}(q^0, q) u_\alpha Q_\beta \bigl[ \epsilon^{\alpha \beta \mu \rho}
    P_T ^{\nu \sigma}  + 
    \epsilon^{\alpha \beta \nu \rho}
    P_T ^{\mu \sigma}  +   ( \rho \leftrightarrow \sigma ) \bigr] ,  
\end{split} 
\end{equation}
where $u_\nu=\delta_0^\nu$ is the velocity of the plasma, and $P_{L,T}$ are two projectors given by 
\begin{equation}
\begin{split}
 P_T ^{\mu \nu} &= \eta^{\mu \nu} - \frac{1}{(u \cdot Q)^2 + Q^2} \left[  
     u \cdot Q  \bigl(u^\mu Q^\nu + u^\nu Q^\nu ) + Q^\mu Q^\nu-Q^2u^\mu u^\nu \right] , \\ 
  P_V^{\mu \nu} &= \eta^{\mu \nu}- \frac{Q^\mu Q^\nu}{Q^2}  - P_\mathbb{T}^{\mu \nu}  .
 \end{split}
\end{equation}
The two scalar functions   $c_{V}(q^0, q)$ and $c_{T}(q^0, q)$ are 
\begin{align}\label{dec} 
c_V(q^0, q) &= 
     \overline{n}_{\nu - \overline{\nu}} \left( \frac{3}{10} Q_1(q^0/q) - \frac{3}{10} Q_3(q^0/q)   \right), \\    
 \label{ten}      
c_T(q^0, q) 
   &= 
     \overline{n}_{\nu - \overline{\nu}}\,  \frac{q^0}{q} 
     \left(- \frac{1}{10} Q_0(q^0/q) + \frac{1}{7} Q_2(q^0/q) - \frac{3}{70} Q_4(q^0/q)  \right) ,     
\end{align}
where $Q_j(x)$ are Legendre functions of the second kind\footnote{In Ref.~\cite{Manes:2012hf} 
the expressions of $c_V$ and $c_T$ were 
written in terms of $Q_1(q^0/q)$ solely, but for our purposes it is advantageous to use this equivalent form.}. 
The  functions above turn out to be the coefficients of the gauge-invariant combinations 
of vector and tensor metric perturbations in Eq.~(\ref{eq:em}).   
In particular, for vector perturbations,  the  asymmetry gives a nonzero contribution to the energy-momentum tensor
 \begin{equation}
 \label{T0i}
 \begin{split}
 \delta\langle T^{0 i}\rangle&= c_V(q^0, q)  \, i \epsilon^{i j k}  q^j (G_k + i q^0 C_k), \\ 
 \delta\langle T^{i j}\rangle&= c_V(q^0, q) \, i q^0  
 \left( \epsilon^{i m n}  \hat{q}^m \hat{q}^j + \epsilon^{j m n}  \hat{q}^m \hat{q}^i  \right)(G_n + i q^0 C_n).
 \end{split}
 \end{equation}
 where $\hat{q}^j=q^j/q$, while    
for tensor perturbations the induced contribution takes the form 
\begin{equation}
 \label{tensor3}
 \delta\langle T^{i j}\rangle= -c_T(q^0, q)\epsilon^{ilm} \delta^{j n}\, i q^l D_{m n}
 +   ( i \leftrightarrow j ) .
 \end{equation}
 In these expressions we have followed the notation of Ref.~\cite{Weinb} for metric perturbations  
 \begin{equation}
 \begin{split}
 h_{0 i} &= G_i ,  \\ 
 h_{i j} &= \frac{\partial C_i}{\partial x^j} + \frac{\partial C_j}{\partial x^i} + D_{i j}, 
 \end{split}
 \end{equation}
 where $G_j(t, \bm{x})$  and $C_j(t, \bm{x})$ are solenoidal vector fields describing  the vector perturbation, 
 and the traceless field $D_{i j}(t, \bm{x})$ satisfying $\partial_i D_{i j}=0$ describes 
 the tensor perturbation.

In order to obtain a kinetic formulation  of these results, it is necessary to introduce the distribution function 
in momentum space and then  derive the Boltzmann equation that governs it. 
The possible contributions to the perturbed energy-momentum tensor from the neutrino asymmetry may be written 
\begin{equation}
\label{perT}
\delta T^{\mu \nu}(t, \bm{x})  = \int  \frac{d^3 p}{p} \, \delta n_{\nu- \overline{\nu}}(\bm{x}, \bm{p}, t) p^\mu  p^\nu, 
\end{equation}  
where $\delta n_{\nu - \overline{\nu}}(t, \bm{x}, \bm{p})$ is the perturbation to the equilibrium neutrino distribution given by 
\begin{equation}
n_{\nu - \overline{\nu}}(t, \bm{x}, \bm{p}) = \overline{n}_+(p) - \overline{n}_-(p) + \delta n_{\nu - \overline{\nu}}(t, \bm{x}, \bm{p}),  
\end{equation}
and $p^\mu = p(1,  \hat{\bm{p}})$. 
A comparison with Eq.(\ref{eq:em}) suggests that 
the integral over $\bm{p}$ in  the Fourier transform of Eq.~(\ref{perT}) may  be viewed as the one-loop integral defining
$-\tfrac{1}{2} \Pi^{\mu \nu \, \rho \sigma}(Q) h_{\rho \sigma}(Q)$. 
The thermal field theory computation shows that,  for $Q \ll |\mu|, T$, 
the corresponding integrand is proportional to $(q^0 -  \hat{\bm{p}} \cdot \bm{q} )^{-1}$. 
Thus,  if in view of (\ref{perT})  we identify it with  $\delta n_{\nu- \overline{\nu}}(Q, \bm{p})$, then we are left with  
\begin{equation}
\label{kin}
(-i q^0 +  i \hat{\bm{p}} \cdot \bm{q} ) \delta n_{\nu- \overline{\nu}}(Q, \bm{p}) = S^{\mu \nu}(Q, \bm{p})h_{\rho \sigma}(Q), 
\end{equation}
where the function $S^{\mu \nu}(Q, \bm{p})$ is determined by the numerator of the one-loop integrand defining 
$c_V$ and $c_T$.
This relation is indeed the Fourier transform of a kinetic equation of the Vlasov type in flat space-time. 
It may be worth pointing out that this connection between the kinetic and  thermal field treatments  is quite common  
within the hard thermal loop approximation. 

Let us now derive the specific form of the kinetic equation corresponding to  Eq.~(\ref{kin}). 
As all the dependence of  the perturbation $\delta n_{\nu- \overline{\nu}}(Q, \bm{p})$ on $p$ is contained  in the 
factor $\bigl(\overline{n}_+' (p) - \overline{n}_-' (p)\bigr)$,  the radial integration over $p$,  including the factor  
$p^\mu p^\nu/p \propto p$,  produces 
the unperturbed distribution $\overline{n}_{\nu - \overline{\nu}}$ of Eqs.~(\ref{dec}) and  (\ref{ten}). 
Thus,  using  a  notation similar to that of  \cite{Weinb}, 
it is convenient to define a 
direction-dependent  
intensity\footnote{The leading perturbation $\delta n_{\nu+ \overline{\nu}}(Q, \bm{p})$ depends on $p$ through the combination 
 $p \bigl(\overline{n}_+' (p) + \overline{n}_-' (p)\bigr)$, 
 and the radial integral of $\delta n_{\nu+ \overline{\nu}}(Q, \bm{p}) p^3$ 
 yields a factor proportional to the unperturbed energy density $\overline{\rho}_{\nu + \overline{\nu}}$.} $K(Q, \hat{\bm{p}})$ through
\begin{equation}
\label{Kdef}
 \overline{n}_{\nu - \overline{\nu}}\,  K(Q, \hat{\bm{p}}) 
\equiv \int_0^\infty \delta n_{\nu- \overline{\nu}}(Q, \bm{p})\ 4 \pi p^3 dp  . 
\end{equation}
In view of Eq.(\ref{perT}), it follows in particular that $K(Q, \hat{\bm{p}})$ must satisfy 
\begin{align}
\overline{n}_{\nu - \overline{\nu}}\int   \frac{d^2 \hat{p}}{4 \pi} \, K(Q, \hat{\bm{p}})\,  \hat{p}_j   &=   \delta T^{0}_{\;\; j}(Q) , \\ 
\label{eq:tenK}
\overline{n}_{\nu - \overline{\nu}}\int   \frac{d^2 \hat{p}}{4 \pi} \, K(Q, \hat{\bm{p}})\,  \hat{p}_i \hat{p}_j    &=   \delta T^{i}_{\;\;  j}(Q)  ,  
\end{align}
where 
$ \delta T^{\mu}_{\;\;\; j}=\delta  T^{\mu j} $ are given  by (\ref{T0i}) and (\ref{tensor3})\footnote{
The spatial indices may be lowered with $\delta_{j k}$,  so that $\hat{p}^i = \hat{p}_i$.}.
To find the kinetic equation for  $K(Q, \hat{\bm{p}})$, we note that the integral 
\begin{equation}
\label{auxvec}
\int \frac{d^2 \hat{p}}{4 \pi}\, \hat{p}_i \hat{p}_n \frac{\hat{\bm{p}} \cdot \bm{q}}{q^0 -  \hat{\bm{p}} \cdot \bm{q} + i 0^+} 
 = A_2 \delta_{i n} + B_2 \hat{q}_i  \hat{q}_n,  
\end{equation}
has the property that  the coefficient $A_2$ is exactly proportional to the one-loop angular integral that yields 
the coefficient $c_V(q^0,q)$:
\begin{equation}
A_2 = 
\frac{q}{2}\int \frac{d^2 \hat{p}}{4 \pi}\, \frac{(1-\hat{\bm{p}} \cdot \hat{\bm{q}}^2) \hat{\bm{p}} \cdot \hat{\bm{q}}}
{q^0 -  \hat{\bm{p}} \cdot \bm{q} + i 0^+} = 
\frac{1}{5}Q_1(q^0/q) - \frac{1}{5}Q_3(q^0/q) .  
\end{equation} 
Hence,  the multiplication of  Eq.~(\ref{auxvec}) by $\epsilon^{n j k} q^j(-a_k + i q^0 F_k)$ yields the same structure proportional to  
$T^{0 i}$ in Eq.~(\ref{T0i}).  
Therefore, we may identify the contribution to $K(Q, \hat{\bm{p}})$ that reproduces 
the effect from the vector perturbation
\begin{equation}
\label{vlasvec}
(-i q^0 + i \hat{\bm{p}} \cdot \bm{q}) K(Q, \hat{\bm{p}}) = 
  \frac{3}{2} \hat{\bm{p}} \cdot \bm{q}\, \hat{p}_n  \epsilon_{n j k}  q_j  \bigl(G_k(Q) + i q^0 C_k(Q)\bigr) . 
\end{equation} 
It can be checked that, upon integration with $\hat{p}_i \hat{p}_j$, this form of $K(Q, \bm{p})$  reproduces  
the correct  $\delta\langle  T^{i j} \rangle$  for the  vector perturbation in (\ref{T0i}). 

We can use a similar argument to find the contribution to $K$ from tensor perturbations. Now the integral 
\begin{equation}
\begin{split}
\int \frac{d^2 \hat{p}}{4 \pi}\, \hat{p}_i \hat{p}_j \hat{p}_l \hat{p}_m \frac{q}{q^0 -  \hat{\bm{p}} \cdot \bm{q} + i 0^+} 
 &= A_4( \delta_{i j}  \delta_{l m} + \text{two similar}) \\ 
 &\quad+  B_4( \delta_{i j}  \hat{q}_l  \hat{q}_m + \text{five similar})  \\ 
 &\quad+  C_4 \, \hat{q}_i  \hat{q}_j \hat{q}_l  \hat{q}_m,  
 \end{split}
\end{equation}
has the coefficient $A_4$ proportional to $c_T(q^0, q)$,  
\begin{equation}
A_4 = 
\frac{q}{8}\int \frac{d^2 \hat{p}}{4 \pi}  \frac{(1-\hat{\bm{p}} \cdot \hat{\bm{q}}^2)^2}{q^0 -  \hat{\bm{p}} \cdot \bm{q} + i 0^+} =
\frac{1}{15} Q_0(q^0/q) -\frac{2}{21} Q_2(q^0/q)+\frac{1}{35} Q_4(q^0/q) , 
\end{equation}
while the others will not contribute to the contraction with $\epsilon q$, 
because of the (anti)symmetry in the indices and the 
transversality property $q_k D_{k n}=0$. 
Including the prefactor $q^0/q$ of (\ref{ten}) and using  (\ref{eq:tenK}), 
one obtains the kinetic equation for  the intensity perturbation that reproduces the parity violating effects 
from tensor fluctuations   
\begin{equation}
\label{vlasten}
(-i q^0 + i \hat{\bm{p}} \cdot \bm{q}) K(Q, \hat{\bm{p}}) = 
  \frac{3}{2} q^0 \hat{p}_i  \epsilon_{i m n}  q_m  \hat{p}_j D_{n j}(Q)  . 
\end{equation} 

The extension of these results to  perturbations of the Robertson-Walker metric can be  made by exploiting the  
invariance under conformal transformations.  
The graviton self-energy is defined by 
\begin{equation}\label{def}
\Pi^{\mu \nu\, \rho \sigma}(x, y) = - 4\left. \frac{\delta \Gamma}{\delta g_{\mu \nu}(x) \delta g_{\rho \sigma}(y)} \right|_{g=\overline{g}} 
 =  -2\left. \frac{ \delta} {\delta g_{\mu \nu}(x)} \left(\sqrt{-g(y)} \langle T^{\rho \sigma}(y)  \rangle\right)\right|_{g=\overline{g}} , 
\end{equation} 
where $\overline{g}$ is a background metric. 
Since the thermal contribution to the underlying  effective action $\Gamma[g_{\mu \nu}]$ is 
conformally invariant, the graviton self-energy for a conformally flat background 
$g_{\mu \nu}(x) = \Omega^2(x) \eta_{\mu \nu}$  
reads 
\begin{equation}
\Pi^{\mu \nu \; \rho \sigma}(x, x')=  \Omega^{-2}(x) \left. \Pi^{\mu \nu \, \rho \sigma}(x-x') \right|_{g=\eta} \Omega^{-2}(x') .  
\end{equation}
As a consequence, the combination $\sqrt{-g(x)} \delta T^{\mu}_{\;\;\; \nu}(x)$ is conformally invariant, 
and  may be evaluated from the already computed $\left. \delta T^{\mu}_{\;\;\; \nu}(x)\right|_{g=\eta}$.  
Therefore,  it is convenient to write the perturbed metric of the expanding universe as
\begin{equation}
ds^2 = \Omega^2(\tau) \bigl(\eta_{\mu \nu}  + h_{\mu \nu}(\tau, \bm{x})\bigr) dx^\mu dx^\nu ,
\end{equation}
where $\tau = \int dt \, a^{-1}(t)$ is the conformal time, and  $\Omega(\tau) = a(t)$. 
By making the replacements  $-i q^0 = \partial_\tau \to a(t) \partial_t$ and $i q_j \to \partial_j$, 
we are left with the kinetic equation for the intensity perturbation $K(t, \bm{x}, \hat{\bm{p}})$, 
\begin{equation}
\label{vlasov}
\frac{\partial K(t, \bm{x}, \hat{\bm{p}})}{\partial t} + \frac{\hat{p}_i}{a(t)} \frac{\partial K(t, \bm{x}, \hat{\bm{p}})}{\partial x^i} = 
\frac{3}{2}\,  \hat{p}_i  \hat{p}_j \epsilon_{i m n} \frac{\partial}{\partial x^m} \left(
\frac{\partial D_{j n}}{\partial t}  +  \frac{\partial^2 C_n}{\partial x^j \partial t} -
\frac{1}{a(t)} \frac{\partial G_n}{\partial x^j}\right) . 
\end{equation}
By assuming that the degeneracy parameter $\xi_\nu$ is preserved in the cosmic expansion, 
the relation (\ref{Kdef}) between the intensity $K(t, \bm{x}, \hat{\bm{p}})$ with  dimensions of energy 
and  $\delta n_{\nu- \overline{\nu}}$ may be written as
\begin{equation}
\label{Kdef2}
 a^3 (t)\,\overline{n}_{\nu - \overline{\nu}}(t) \,  K(t, \bm{x}, \hat{\bm{p}}) 
=  \int_0^\infty \delta n_{\nu- \overline{\nu}}(t, \bm{x}, \bm{p})\ 4 \pi p^3 dp  ,  
\end{equation}
where  the fermion asymmetry $\overline{n}_{\nu - \overline{\nu}}(t)  \equiv  \overline{T}^3(t) \xi_\nu/6$
has now been expressed in terms of the equilibrium temperature $\overline{T}(t) = T_0/a(t)$ in the comoving system. 
With this relation and Eq.~(\ref{vlasov}),  one obtains the  Boltzmann equation for the perturbation 
$\delta n_{\nu - \overline{\nu}}(t, \bm{x}, \bm{p})$:  
\begin{equation}
\label{vlasov2}
\begin{split}
\frac{\partial \delta n_{\nu - \overline{\nu}}(t, \bm{x}, \bm{p})}{\partial t} + 
\frac{\hat{p}_i}{a(t)} \frac{\partial \delta n_{\nu - \overline{\nu}}(t, \bm{x}, \bm{p})}{\partial x^i} &= 
-\frac{1}{2} \bigl(\overline{n}_{+}'(p) - \overline{n}_{-}'(p)\bigr) \\ 
& \quad \times  \hat{p}_i  \hat{p}_j \epsilon_{i m n} \frac{\partial}{\partial x^m} \left(
\frac{\partial D_{j n}}{\partial t}  +  \frac{\partial^2 C_n}{\partial x^j \partial t} - 
\frac{1}{a(t)} \frac{\partial G_n}{\partial x^j}\right) .   
\end{split}
\end{equation}
 To determine the components  $\delta T^{0}_{\;\;\; j}(t, \bm{x})$ and  $\delta T^{i}_{\;\; j}(t, \bm{x})$ in the  usual comoving coordinates, 
we can use the relations 
\begin{align}
\label{rela1}
\Omega^4(\tau) \delta T^{\tau}_{\;\;  j}(\tau, \bm{x}) &= a^3(t) \delta T^{0}_{\;\; j}(t, \bm{x}) = 
\left. \delta T^{\tau}_{\;\; j}(\tau, \bm{x)}\right|_{g=\eta} , \\ 
\Omega^4(\tau) \delta T^{i}_{\;\; j}(\tau, \bm{x}) &= a^4(t) \delta T^{i}_{\;\; j}(t, \bm{x}) = 
\left. \delta T^{i}_{\;\; j}(\tau, \bm{x)}\right|_{g=\eta} . 
\end{align}
which lead to 
\begin{align}
\label{deltaT}
 \delta  T^{0}_{\;\;j}(t, \bm{x}) &=  \overline{n}_{\nu - \overline{\nu}}(t) \int   \frac{d^2 \hat{p}}{4 \pi} \, K(t, \bm{x}, \hat{\bm{p}})\,  \hat{p}_j   , \\ 
\label{tenK}
 \delta  T^{i}_{\;\; j}(t, \bm{x}) &= 
 \frac{\overline{n}_{\nu - \overline{\nu}}(t)}{a(t)} \int   \frac{d^2 \hat{p}}{4 \pi} \, K(t, \bm{x}, \hat{\bm{p}})\,  \hat{p}_i \hat{p}_j    .  
\end{align}

The simplicity of the source terms in Eq.~(\ref{vlasov}) or (\ref{vlasov2}) is remarkable. 
A nice feature of this result is that, in the absence of $G_n$,  
the effect of the coefficients $c_V$ and $c_T$ in the kinetic equation enter through   
the single combination  of vector and tensor quantities corresponding to 
the spatial perturbation of the metric,  $a^{-2} \delta g_{i j} =D_{i j} + \partial_j C_i +  \partial_i C_j$.
This is similar to what happens  in the Boltzmann equation~\cite{Rebhan:1994zw,Weinberg:2004fk}
 for the leading even-parity density perturbation 
$\delta n_{\nu + \overline{\nu}}(t, \bm{x}, \bm{p})$: 
\begin{equation}
\label{vlasov3}
\begin{split}
\frac{\partial \delta n_{\nu + \overline{\nu}}(t, \bm{x}, \bm{p})}{\partial t} + 
\frac{\hat{p}_i}{a(t)} \frac{\partial \delta n_{\nu + \overline{\nu}}(t, \bm{x}, \bm{p})}{\partial x^i} &= 
\frac{1}{2} \, p \bigl(\overline{n}_{+}'(p) + \overline{n}_{-}'(p)\bigr) \\ 
& \quad \times  \hat{p}_j  \hat{p}_n   \frac{\partial}{\partial t}\left(
D_{j n} +  \frac{\partial C_n}{\partial x^j} + \frac{\partial C_j}{\partial x^n}\right) . 
\end{split}
\end{equation}
 The relation between the field theory approach and the one based on kinetic theory has been recently established  
in \cite{Son:2012zy},  where the authors have considered the effects of triangle anomalies without any metric perturbation. 
The previous treatment completes the derivation of the kinetic equation 
by including parity violation effects from chiral matter 
in the presence of  a weak time-dependent gravitational field.

It is instructive to write the explicit form of the odd parity corrections to the anisotropic inertia and to compare them with 
the leading contributions proportional to the energy-density $\overline{\rho}_{\nu + \overline{\nu}}(t)$.  
Here we reproduce for convenience the governing equations for these quantities~\cite{Weinb}:  
\begin{align}
\label{vlasov4}
\frac{\partial J(t, \bm{x}, \hat{\bm{p}})}{\partial t} + \frac{\hat{p}_i}{a(t)} \frac{\partial J(t, \bm{x}, \hat{\bm{p}})}{\partial x^i} &= 
-2 \hat{p}_j  \hat{p}_n
\left(
\frac{\partial D_{j n} }{\partial t} - \frac{2}{a(t)} \frac{\partial \tilde{G}_j}{\partial x^n} \right) ,  \\  
\label{deltaTd}
 \delta  T^{0}_{\;\;j}(t, \bm{x}) &=  a(t) \overline{\rho}_{\nu + \overline{\nu}}(t) \int   \frac{d^2 \hat{p}}{4 \pi} \, J(t, \bm{x}, \hat{\bm{p}})\,  \hat{p}_j   , \\ 
\label{tenKd}
 \delta  T^{i}_{\;\; j}(t, \bm{x}) &= 
 \overline{\rho}_{\nu + \overline{\nu}}(t) \int   \frac{d^2 \hat{p}}{4 \pi} \, J(t, \bm{x}, \hat{\bm{p}})\,  \hat{p}_i \hat{p}_j  ,  
\end{align}
where  $\tilde{G}_j \equiv G_j  - a \partial_t C_j$. 
In order to find the time dependence of $\delta  T^{\mu}_{\;\;\;\nu}$, we could use Eqs.~(\ref{T0i}) and (\ref{tensor3}), 
and  evaluate the inverse Fourier transforms. 
But it is better to integrate the Vlasov equations,  
and then  compute (\ref{deltaT}) and (\ref{deltaTd}),
because the initial conditions are more clearly introduced in this way.
With the  standard expansion in plane waves $e^{i \bm{q} \cdot \bm{x}}$, 
this procedure yields the time dependence of $K(t,\bm{q}, \hat{p})$ and $J(t,\bm{q}, \hat{p})$, which upon evaluation of the integrals in  
(\ref{deltaT}) and (\ref{deltaTd}) for a vector perturbation yields
\begin{equation}
\begin{split}
\delta  T^{0}_{\;\;j}(t, \bm{q}) &= \tilde{g}_j(t, \bm{q}) + 4 a(t) \overline{\rho}_{\nu + \overline{\nu}}(t)  
  \int_{0}^u \, \frac{j_2(u - u')}{u-u'}  \tilde{G}_j(t', \bm{q}) du'  \\ 
&\quad 
-\frac{3}{2} \overline{n}_{\nu - \overline{\nu}}(t)  q 
  \int_{0}^u \, \frac{j_2(u - u')}{u-u'} \epsilon_{j m n} i \hat{q}_m \tilde{G}_n(t', \bm{q}) du'  , 
\end{split}
\end{equation}
where $u$ is proportional to  the conformal time, 
\begin{equation}
u = q \int_{t_1}^t \frac{dt'}{a(t')} ,
\end{equation}
and $\tilde{g}_i(t, \bm{q})$ is any arbitrary invariant contribution satisfying  $q_j \tilde{g}_j = 0$.   
This may be traced to the 
 solution 
of the Vlasov equation in the absence of sources for a specific initial condition $J(t_1, \bm{q}, \hat{\bm{p}})$, 
\begin{equation}
\tilde{g}_i(t, \bm{q}) = a(t) \int \frac{d^2 \hat{p}}{4 \pi} \, \exp \left(-i \hat{\bm{p}} \cdot \bm{q} 
  \int_{t_1}^{t} \frac{dt' }{a(t')}  \right) J(t_1, \bm{q}, \hat{\bm{p}}) \hat{p}_i . 
\end{equation}
In the case of tensor modes the total contribution to $\delta  T^{k}_{\;\;j}$ reads
\begin{equation}
\begin{split}
\delta  T^{k}_{\;\;j}(t, \bm{q}) &=\tilde{d}_{k j}(t, \bm{q}) - 4 \overline{\rho}_{\nu + \overline{\nu}}(t)  
  \int_{0}^u \, \frac{j_2(u - u')}{(u-u')^2}  \frac{\partial D_{k j}(t', \bm{q})}{\partial t'} dt'  \\ 
&\quad +  
\frac{3}{2} \frac{\overline{n}_{\nu - \overline{\nu}}(t) q}{a(t)}  
  \int_{0}^u \, \frac{j_2(u - u')}{(u-u')^2} \left( \epsilon_{k m n} i \hat{q}_m 
  \frac{\partial D_{j n}(t', \bm{q})}{\partial t'} + ( k \leftrightarrow j ) \right) dt'  , 
\end{split}
\end{equation}
where the traceless divergenceless part $\tilde{d}_{ kj}(t, \bm{q})$ plays the same role as before.  
The kernels with spherical Bessel function arise from the integrals
\begin{equation}
\begin{split}
\int   \frac{d^2 \hat{p}}{4 \pi}  e^{-i \hat{\bm{p}} \cdot \hat{\bm{q}}\, u} \, \hat{\bm{p}} \cdot \hat{\bm{q}} \, 
\hat{p}_i \hat{p}_j &= -i \frac{j_2(u)}{u}  \delta_{i j} + \ldots\\ 
\int   \frac{d^2 \hat{p}}{4 \pi}  e^{-i \hat{\bm{p}} \cdot \hat{\bm{q}}\, u} \, 
\hat{p}_i \hat{p}_j \hat{p}_l \hat{p}_m&= \frac{j_2(u)}{u^2} ( \delta_{i j}  \delta_{l m} + \text{two similar}) + \ldots , 
\end{split}
\end{equation} 
and, as expected, they exactly  agree with the inverse Fourier transform of $c_V$ and $c_T$, 
\begin{equation}
\begin{split}
\int_{-\infty + i 0^+}^{\infty + i 0^+} \frac{d q^0}{2 \pi} e^{- i q^0 \tau}
\left(\frac{3}{10} Q_1(q^0/q) - \frac{3}{10} Q_3(q^0/q) \right) & = -\frac{3}{2} \frac{j_2(q \tau)}{\tau} \Theta(\tau), \\
\int_{-\infty + i 0^+}^{\infty + i 0^+} \frac{d q^0}{2 \pi} e^{- i q^0 \tau}
\left(-\frac{1}{10} Q_0(q^0/q) + \frac{1}{7} Q_2(q^0/q) - \frac{3}{70} Q_4(q^0/q) \right) &=  \frac{3i }{2 q} \frac{j_2(q \tau)}{\tau^2} \Theta(\tau). 
\end{split}
\end{equation}

To conclude, let us consider in more detail the equation for the tensor modes. If we choose the polarization tensors 
$e_{j n}(\hat{\bm{q}}, \lambda)$ to be  the ones  produced by the rotation  which takes    
$\hat{\bm{z}} \to \hat{\bm{q}}$, where 
\begin{equation}
e_{j n}(\hat{\bm{z}}, \lambda =  \pm 2) = 
\begin{pmatrix}
\frac{1}{\sqrt{2}} & \frac{i \lambda}{2 \sqrt{2}} & 0 \\ 
\frac{i \lambda}{2 \sqrt{2}} & -\frac{1}{\sqrt{2}} & 0 \\ 
0 & 0 & 0  
\end{pmatrix} ,
\end{equation} 
one can easily check the identity 
\begin{equation}
\epsilon_{k m n} i \hat{q}_m  e_{j n}(\bm{q}, \lambda) + \epsilon_{j m n} i \hat{q}_m  e_{k n}(\bm{q}, \lambda)  = 
  \lambda\, e_{k j}(\bm{q}, \lambda) ,
\end{equation}
Thus the decomposition of the tensor modes according to 
\begin{equation}
D_{j n}(t, \bm{q}) = \sum_\lambda e_{j n}(\bm{q}, \lambda) \mathcal{D}(t, \bm{q}, \lambda), 
\end{equation}
leads to decoupled equations for the quantities $\mathcal{D}(t, \bm{q}, \lambda)$. 
In the absence of $\tilde{d}_{j n}$, the Einstein equations adopt the form 
\begin{equation}
\begin{split}
16 \pi G \biggl(- 4 \overline{\rho}_{\nu + \overline{\nu}}(t)  & +  
\frac{3 \lambda}{2} \frac{\overline{n}_{\nu - \overline{\nu}}(t) q}{a(t)} \biggr)
  \int_{0}^u \, \frac{j_2(u - u')}{(u-u')^2}  \frac{\partial \mathcal{D}(t', \bm{q}, \lambda)}{\partial t'} dt'  \\   
&\quad= \frac{\partial^2 \mathcal{D}}{\partial t^2} + \frac{3\dot{a}(t)}{a(t)} \frac{\partial \mathcal{D}}{\partial t} + \frac{q^2}{a^2} \mathcal{D}. 
\end{split}
\end{equation} 
Due to the non-zero net neutrino number, these equations are not longer independent of the helicity $\lambda$,   
but for each helicity the equation has the same previously known form~\cite{Weinberg:2004fk}, and  
the same techniques may be used to find solutions~\cite{Dicus:2005fk}. 
The main effect of  $\overline{n}_{\nu - \overline{\nu}}$ is to produce birefringence or a splitting of the two helicities, which increases linearly with $q$. 
The relative size of this correction is therefore wave number-dependent,  
$\overline{n}_{\nu - \overline{\nu}}\,  q /\overline{\rho} a \sim \xi_\nu q/\overline{T} a$. 
Whether  this has a non-negligible impact on the spectrum of primordial gravity waves is an issue to be further considered.

\begin{acknowledgments}
I am grateful to Jos\'e J. Blanco-Pillado for his comments and Juan L. Ma\~nes for a careful reading of the manuscript.
This research was supported in part by the Spanish Ministry of Science and Technology under Grants FPA2009-10612 and FPA2012-34456, 
the Spanish Consolider-Ingenio 2010 Programme CPAN (CSD2007-00042), and by the Basque Government under Grant IT559-10.
\end{acknowledgments}

\bibliography{bibasymm}

\end{document}